\documentclass[aps,prl,twocolumn,groupedaddress]{revtex4}
\usepackage{graphicx}
\begin{document}

\title{Holographic Shell Model: Stack Data Structure inside Black Holes}

\author{Aharon Davidson}
\email[Email: ]{davidson@bgu.ac.il}
\homepage[~Homepage: ]{http://www.bgu.ac.il/~davidson/}

\affiliation{Physics Department, Ben-Gurion University,
Beer-Sheva 84105, Israel}

\date{June9, 2011}

\begin{abstract}
We suggest that bits of information inhabit, universally and holographically,
the entire black hole interior, a bit per a light sheet unit interval of order
Planck area difference.
The number of distinguishable (tagged by a binary code) configurations,
counted within the context of a discrete holographic shell model, is given
by the Catalan series.
The area entropy formula is recovered, including the universal logarithmic
correction, and the equipartition of mass per degree of freedom
is proven.
The black hole information storage resembles a stack data structure.
\end{abstract}

\pacs{}

\maketitle
The 't Hooft-Susskind holographic principle\cite{Hprinciple} asserts
that all of information contained in some region of space can be
represented as a 'hologram' on the boundary of that region.
It furthermore puts a universal purely geometrical bound, saturated
by Bekenstein-Hawking area entropy formula \cite{Bekenstein-Hawking},
on the amount of entropy stored within that region, namely
\begin{equation}
	S\leq \frac{A}{4G}~.
\end{equation}
Here, $A$ denotes the area of the closed spacial boundary, $G$
is Newton's constant, and $\hbar=c=k_B=1$.
Sticking momentarily to spherical symmetry, a covariant \cite{Bousso} 
local generalization of the holographic bound, that is
\begin{equation}
	S(r)\leq \frac{\pi r^2}{G}~,
	\label{Sr}
\end{equation}
is expected to hold for every concentric sphere of circumferential radius $r$.
This, unfortunately, seems to fall short within the inner region of a black hole.
As far as entropy packing is concerned, the interior of a black
hole is apparently superfluous.
The first local realization of maximal entropy packing has been
demonstrated \cite{DavidsonGurwich} within the context of spontaneously
induced ($G^{-1}$ treated as a VEV) general relativity.
The black hole limit is then governed by a phase transition
\cite{phase} which occurs precisely at the would have been horizon.
The recovered exterior Schwarzschild solution of mass $M$ connects
with a novel core of vanishing spatial volume characterized by the
non-singular Komar mass distribution
\begin{equation}
	M(r)=\frac{r^2}{4G^2 M}~.
	\label{Mr}
\end{equation}
The emerging onion-like maximal entropy packing profile is such that the
entropy $S(r)$ of any inner sphere is unaffected by the outer layers; any
additional incoming entropy is maximally packed on its own external layer.
In some sense, each layer acts as a stretched horizon
\cite{stretch,Padmanabhan}.
Spontaneously induced general relativity sheds new light on the way information
is stored within a black hole.
Rather than envision bits of information evenly spread on the horizon surface,
they now inhabit, universally and holographically, the entire black hole interior.

Within such a core, consider a light sheet interval \cite{lightsheet} constructed
as the difference between two concentric spherical cones (outer radius $r_{out}$,
inner radius $r_{in}$) sharing a common solid angle $\Omega\leq 4\pi$.
A light sheet \emph{unit} interval is then defined to have order Planck area
difference, that is
\begin{equation}
	\Omega\left( r_{out}^2-r_{in}^2\right )=\eta\ell_{Pl}^2~.
	\label{unit}
\end{equation}
The ${\cal O}(1)$ coefficient $\eta$ is regarded fundamental, and will be
determined later.
In this paper, rather than tiling the horizon by $n$ Planck  area patches, the
traditional way, we suggest \emph{filling up the interior with $n$ light sheet
unit intervals.}
It is useful, but not mandatory, to think of a specific shape of these intervals;
it is only their standard size and relative localization which are relevant.
This way, we do not need to further assume several microstates per bit,
but only count the total number $g_n$ of distinguishable spatial arrangements
of $n$ indistinguishable degrees of freedom.
At the fundamental level, the counting calls for graph enumeration which is
currently unknown for large $n$.
Consequently, recalling the discrete black hole prototype
\cite{Bekenstein,BekensteinMukhanov}, we construct a discrete holographic
shell model, where the number of distinguishable configurations (tagged
by a Wheeler's 'it from bit' binary code) is given by the Catalan series.
While regarding ref.{\cite{DavidsonGurwich}} to serve as the underlying
rationale, our model may have life of its on.
Discrete black holes have been conjectured \cite{lab} to be created in
the lab, and the possibility of detecting terrestrial mini black holes has been
recently raised \cite{earth}. 

By definition, a shell model assumes the arrangement of $n$ degrees of freedom
in concentric spherical shells, such that the $i$-th shell (counting from the origin
outwards) hosts $k_i$ degrees of freedom, with $\sum k_i=n$.
Put in a different way, the exterior shell which hosts $k$ degrees of freedom
surrounds a core which exhibits a full shell structure for the residual $(n-k)$
degrees of freedom.
All $g_n$ different configurations are then being counted, without replication,
by letting $k$ run from $1$ to $n$.
Denoting by $f_k$ the total number of distinguishable configurations of $k$
degrees of freedom within a given shell, with $f_k<g_k$ (and $f_0\leq g_0$)
on consistency grounds, we are led to the recursion formula
\begin{equation}
	f_0 g_{n}=\sum_{k=1}^{n}f_k~g_{n-k}~,
	\label{recursion}
\end{equation}
which holds for $n=1,2,..$.
Notice that, reflecting our universal approach, this formula does not necessarily
depend on the total number of spacetime dimensions. 
The formal solution $g_n (f_1,...,f_n)$ of eq.\ref{recursion} is given by
\begin{equation}
	\frac{g_n}{g_0}=\sum_{i,j}
	\frac{(j_1+...+j_\ell)!}{j_1!...j_\ell !}
	\frac{{f_{i_1}}^{j_1}...{f_{i_\ell}}^{j_\ell}}{{f_0}^{j_1+...+j_\ell}}~,
\end{equation}
where the positive integers $i_p,j_p$ are subject to the Diophantine equation
$\sum_{p} i_p j_p=n$.
Obviously, there is no reason for the respective scales $g_0$ and $f_0$ to
differ from each other, so we set $f_0=g_0$ from this point on.
Furthermore, we find it quite tempting, and presumably even required, to
assign $f_0= g_0=1$ for 'nothing'. 

One must be more specific with regards to what defines a shell in the first
place, at least at the algebraic level, by fixing or restricting the $f_k$ series.
Adopting the working principle that one and only one information storing
mechanism exists, our discussion takes two paths.
%Either the $f_k$ series is trivial or else it is not too different from the $g_k$
%series itself.

\begin{figure}[th]
	\includegraphics[scale=0.6]{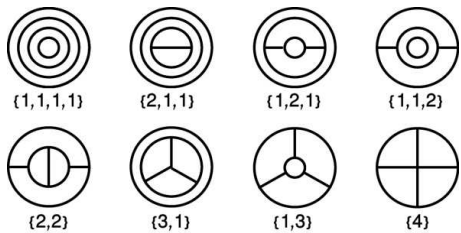}
	\caption{A geometrical representation of the distinguishable
	configurations for the simple model is demonstrated for $n=4$
	(notice the equal length difference intervals).
	The shell structure of each configuration is specified.}
	\label{1}
\end{figure}

\textbf{\emph{A simple model:}}
First, consider the possibility that the inner structure of any given shell
is trivial.
This triviality can be translated into
\begin{equation}
	f_k=f_0\quad \Longrightarrow \quad g_n=2^{n-1}g_0~,
	\label{simple}
\end{equation}
giving rise to the statistical entropy
\begin{equation}
	S_n=(n-1)\log 2+\log g_0 
	\quad\quad (n=1,2,...)~.
	\label{Ssimple}
\end{equation}
The unfortunate feature that $g_0$ is not the zeroth element of the $g_n$
series can be considered a drawback, telling us that the choice $g_0=1$
cannot be enforced in this case.
In fact, such prototype discrete black hole models, either with $g_0=1$, or
alternatively with $g_0=2$, have been discussed by Bekenstein and
Mukhanov \cite{BekensteinMukhanov} and by many others \cite{discrete},
although not in the context of a shell model.
A geometric representation of the simple shell structure is depicted in
Fig.\ref{1}.
%%% Fig.1 %%%

\textbf{\emph{An elaborate model:}}
A non-trivial $f_k$ must obey $f_k<g_k$, and is expected to exhibit a
$g_k$-like structure.
The simplest tenable choice is
\begin{equation}
	f_k=g_{k-1}~,
	\label{sophisticated}
\end{equation}
suggesting that it actually takes one degree of freedom to close a shell.
In other words, resembling a Russian nested doll, each shell exhibits
an inner shell structure, and so on.
The emerging refined recursion formula, with eq.(\ref{sophisticated})
substituted into eq.(\ref{recursion}), is quite familiar and is known to
generate the famous Catalan series; this series is given explicitly by
$C_n=1,1,2,5,14,42,132,..$.
To be more specific, the total number of configurations is then
\begin{equation}
	g_n=\frac{(2n)!}{(n+1)!n!}g_0 \quad\quad  (n=0,1,2,...)~,
\end{equation}
and by means of Stirling's formula, the large-$n$ entropy
expansion takes the form
\begin{equation}
	S_n \simeq n\log 4-\frac{3}{2}\log n+\log\frac{g_0}{\sqrt{\pi}} +...~.
	\label{Sn}
\end{equation}
Unlike in the simple model, this time $g_0$ does constitute the zeroth
element of the $g_n$ series.
We may thus assign $g_0=1$ to the vacuum state to end up with $S_1=0$,
and arrive at the non trivial yet far reaching conclusion that
the associated black hole ground state is non-degenerate.
 
The emergence of the logarithmic term in the entropy expression eq.\ref{Sn},
with the particular coefficient $-\frac{3}{2}$, is by no means trivial.
With the Cardy formula \cite {Cardy} as a field theoretical starting point,
first-order corrections to the Bekenstein-Hawking entropy have been calculated
\cite{Carlip} for a variety of theories.
Despite very different physical assumptions, these corrections are always
proportional, with the same $-\frac{3}{2}$ factor, to the logarithm of the horizon
surface area.
The field theoretical corrections turn out to be in agreement with the
Kaul-Majumdar \cite{geometricS} 'quantum geometrical' approach, and
even with loop gravity models \cite{loop}.
From any direction viewed, the universality of the $-\frac{3}{2}$ coefficient,
shared by our elaborate model (eq.\ref{Sn}), is manifest and still very
mysterious.

\begin{figure}[th]
	\includegraphics[scale=0.65]{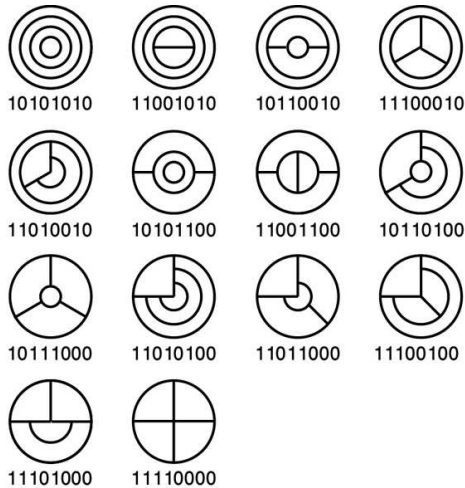}
	\caption{A geometrical representation of the Catalan configurations is
	demonstrated for $n=4$ (notice that the equal area difference intervals
	are represented by equal length difference intervals).
	Each configuration is tagged by a Wheeler style binary code.}
	\label{2}
\end{figure}

\begin{figure}[th]
	\includegraphics[scale=0.6]{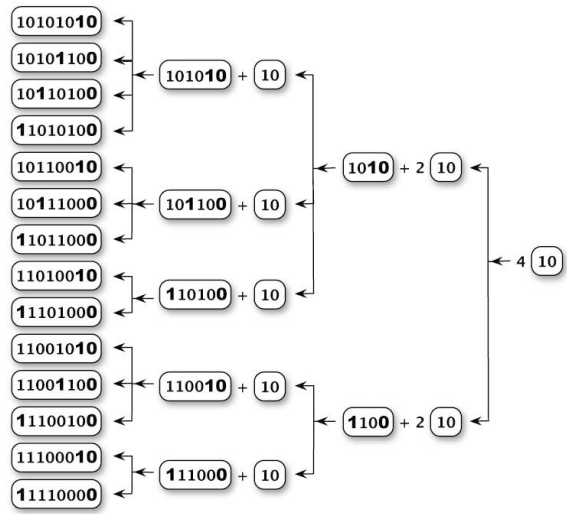}
	\caption{The construction tree of an order-$n$ black hole from $n$
	fundamental black holes is sketched for $n=4$.
	The $1,0$ digits associated with the last-in-first-out degree of
	freedom are highlighted.}
	\label{3}
\end{figure}

There are many counting problems in combinatorics whose solution is given
by the Catalan numbers, two of which are relevant to the present discussion.
For example, $C_n$ counts the number of expressions containing n pairs of
parentheses which are correctly matched.
In this language, $\{\}$ is a fundamental building block, irreducible
$\{\cdots\}$ stands for a single shell (with $\cdots$ denoting its inner shell
structure), so that $\{\cdots\},...,\{\cdots\}$ represents a full shell model
configuration.
Reading this configuration as a mathematical expression from left to right is
equivalent to moving from the inside outwards.
Adding a degree of freedom to a given configuration can be schematically
described by closing from the r.h.s. by $\}$, and inserting $\{$ at some
allowed inner location.
Using the dictionary $\{= 1$ and $\}=0$, every distinguishable configuration
is then tagged by a binary code which captures its construction from individual
indistinguishable degrees of freedom.
This constitutes a realization of Wheeler's 'it from bit' philosophy; a missing
ingredient, perhaps, is the rule which governs the transitions among the various
configurations.
A geometric representation of the distinguishable Catalan configurations
(demonstrated for $n=4$) is depicted in Fig.\ref{2} (notice that the equal area
difference intervals are represented by equal length difference intervals).
Evidently, the set of configurations of the simple model constitutes a subset
of the Catalan configurations.
%%% Fig.2 %%%
Altogether, see Fig.\ref{3}, there are exactly $C_n$ ordered roots
to construct an order-$n$ black hole by adding, step by step, fundamental
order-$1$ black holes.

In computer science, a data structure is a particular way of storing and
organizing data so that it can be used efficiently.
The Catalan numbers happen to be directly related to a well known restricted
data structure called a 'stack', thereby suggesting that even Nature's
ultimate information storage, a black hole, exhibits a text book standard data
structure.
A stack is nothing but a last-in-first-out (LIFO) data structure whose limited
number of operations solely consists of 'push' ($\equiv$ open parentheses) and
'pop' ($\equiv$ close parentheses). 
$C_{n}$ is then the number of stack-sortable permutations of the ordered list
of integers $1, 2, ... , n$.
Beware that the analogy is solely in the count procedure, and information does
not really exit the black hole the way it does from a standard data structure.
%%% Fig.3 %%%

A direct consequence of filling up the entire black hole interior by an arbitrary
integer number of light sheet unit intervals (define by eq.\ref{unit}) is the area
quantization
\begin{equation}
	A_n=n A_1~.
	\label{An}
\end{equation}
This is fully consistent with the apparent role of the horizon surface area
as a quantum mechanical adiabatic invariant \cite{Bekenstein}, where an
optional constant term has been dropped off to assure $A_{0}=0$.
The consistency of eq.\ref{An} is also reflected by the fact that by its
substitution into the modified (logarithm corrections included) asymptotic
area entropy formula derived by a variety of authors
\cite{Cardy,Carlip,geometricS,loop}, eq.\ref{Sn} gets forcefully recovered.

Only our simple model is capable of literally reconciling with the
Bekenstein-Hawking black hole entropy formula.
Eq.\ref{Ssimple} and eq.\ref{An} can coincide, with
\begin{equation}
	A_1=4 \ell_{Pl}^2 \log2, ~~ S_{1}=\log 2,
	~~ M_1= \sqrt{\frac{\log 2}{\pi}}M_{Pl}~,
\end{equation}
provided one is ready to accept a doubly degenerate ground state, that is
$g_1=2$.
The non-degenerate case $g_1=1$, for which $S_{1}=0$, may also fit
in, but it requires supplementing the Bekenstein-Hawking area entropy by a
constant term, which is of course legitimate. 

New doors get open, however, in case the Bekenstein-Hawking
entropy formula happens to be, albeit mandatory, just a large-$n$ limit.
In particular, as far as our elaborate model is concerned, we do asymptotically
recover the equal spacing feature $\Delta S=2\log 2$ up to
${\cal O}(\frac{1}{n})$ corrections. 
This allows us to calculate the $A_1$ area coefficient, so that the
fundamental ground state black hole is now associated with
\begin{equation}
	A_1=8\ell_{Pl}^2 \log2, ~~ S_1=0,
	~~ M_1= \sqrt{\frac{\log 2}{2\pi}}M_{Pl}~,
\end{equation}
and thereby determine the fundamental coefficient $\eta$, see eq.{\ref{unit},
to be $8\log 2$.
Notice that the area is now twice as large in comparison with the
simple model.

The mass distribution which accompanies the holographic shell model is in
some respect counter intuitive.
It can be easily verified that the overall mass of an order-$n$ black hole is
given by
\begin{equation}
	M_n=\frac{1}{4G}\sqrt{\frac{n A_1}{\pi}}~.
\end{equation}
But what about the partial mass $M_{k, n}$ associated with the order-$k$
inner concentric section which resides inside a larger order-$n$ black hole?
Recalling that $S_{k,n}=S_k $, a fact which expresses the universal nature of the
holographic entropy packing, one could have naively expected $M_{k,n}=M_k$,
but this is certainly not the case.
A closer inspection, based on eq.\ref{Mr}, reveals that
\begin{eqnarray}
	&&M_{k, n}=\frac{k}{n}M_n=\sqrt{\frac{k}{n}} M_k < M_k~,\\
	&&\frac{S_{k,n}}{M_{k,n}}=
	\sqrt{\frac{n}{k}}\frac{S_k}{M_k}>\frac{S_k}{M_k}~.
\end{eqnarray}
The conclusion is threefold:

\noindent (i)  The mass per degree of freedom, namely $M_n/n$, depends
 on the black hole size.
It costs much less gravitational energy to store one degree of freedom inside
a large black hole than inside a small black hole.

\noindent (ii) For a given $n$, however, the allocation of mass per degree of
freedom, that is $M_{k,n}/k$, is $k$-independent.
All degrees of freedom are worth the same, a true equipartition, no matter where
they are located.
This pleasing result accounts for the self consistency of the shell model.

\noindent (iii) 
Reflecting the $\sqrt{\frac{n}{k}}>1$ factor, the entropy to energy ratio calculated
for the order-$k$ inner core section is in apparent violation of Bekenstein's universal
entropy bound \cite{Bbound}.
It is thus important to note that Bekenstein's bound is relevant only for weakly
self gravitating isolated physical systems, which is not the case here, and for
these it is a much stronger bound than the holographic one.

To summarize, the discrete holographic shell model has been carefully designed
to capture the holographic entropy packing inside black holes.
In this model, the light sheet unit interval has been elevated to the level of
the fundamental geometrical building block.
This way, we conceptually deviate from the traditional dogma of tiling the
black hole horizon by Planck area patches.
By counting the total number of distinguishable configurations, and tagging
them with a binary code, the Catalan series has made
a remarkable entrance into black hole physics.
The area entropy formula has been recovered, including in particular its universal
logarithmic correction, and the equipartition of mass per degree of freedom proven.
The conclusion that Nature's ultimate information storage resembles a text
book standard data structure sheds new light on the black
hole information puzzle.

\acknowledgments{}
It is a pleasure to thank David Owen  for constructive comments, and
Ilya Gurwich, Shimon Rubin, and Ben Yellin for their valuable feedback. 
Special thanks to BGU president Prof. Rivka Carmi for the kind support.

\end{document}